\documentclass[%
 aip,
 amsmath,amssymb,
 reprint,%
]{revtex4-1}

\usepackage{graphicx}
\usepackage{dcolumn}
\usepackage{bm}

\usepackage[utf8]{inputenc}
\usepackage[T1]{fontenc}
\usepackage{mathptmx}
\usepackage{physics}

\begin{document}

\preprint{AIP/123-QED}

\title[]{Few-electrode design for silicon MOS quantum dots}

\author{Eduardo B. Ramirez}

\affiliation{Institute for Quantum Computing, University of Waterloo, Waterloo, Ontario, Canada N2L 3G1}
\affiliation{Department of Physics and Astronomy, University of Waterloo, Waterloo, Ontario, Canada, N2L 3G1}

\author{Francois Sfigakis}

\affiliation{Institute for Quantum Computing, University of Waterloo, Waterloo, Ontario, Canada N2L 3G1}
\affiliation{Department of Chemistry, University of Waterloo, Waterloo, Ontario, Canada, N2L 3G1}

\author{Sukanya Kudva}

\affiliation{Institute for Quantum Computing, University of Waterloo, Waterloo, Ontario, Canada N2L 3G1}
\affiliation{Indian Institute of Technology Bombay, Mumbai, India 400076}

\author{Jonathan Baugh}
\email{baugh@uwaterloo.ca}
\affiliation{Institute for Quantum Computing, University of Waterloo, Waterloo, Ontario, Canada N2L 3G1}
\affiliation{Department of Chemistry, University of Waterloo, Waterloo, Ontario, Canada, N2L 3G1}

\date{\today}

\begin{abstract}
Silicon metal-oxide-semiconductor (MOS) spin qubits have become a promising platform for quantum information processing, with recent demonstrations of high-fidelity single and two-qubit gates. To move beyond a few qubits, however, more scalable designs that reduce the fabrication complexity and electrode density are needed. Here, we introduce a two-metal-layer MOS quantum dot device in which tunnel barriers are naturally formed by gaps between electrodes and controlled by adjacent accumulation gates. The accumulation gates define the electron reservoirs and provide tunability of the tunnel rate of nearly 8.5 decades/V, determined by a combination of charge sensor electron counting measurements and by direct transport. The valley splitting in the few-electron regime is probed by magneto-spectroscopy up to a field of 6 T, providing an estimate for the ground-state gap of 290 $\mu$eV. We show preliminary characterization of a double quantum dot, demonstrating that this design can be extended to linear dot arrays that should be useful in applications like electron shuttling. These results motivate further innovations in MOS quantum dot design that can improve the scalability prospects for spin qubits. 
\end{abstract}

\maketitle
\begin{figure}
\includegraphics[scale=0.09]{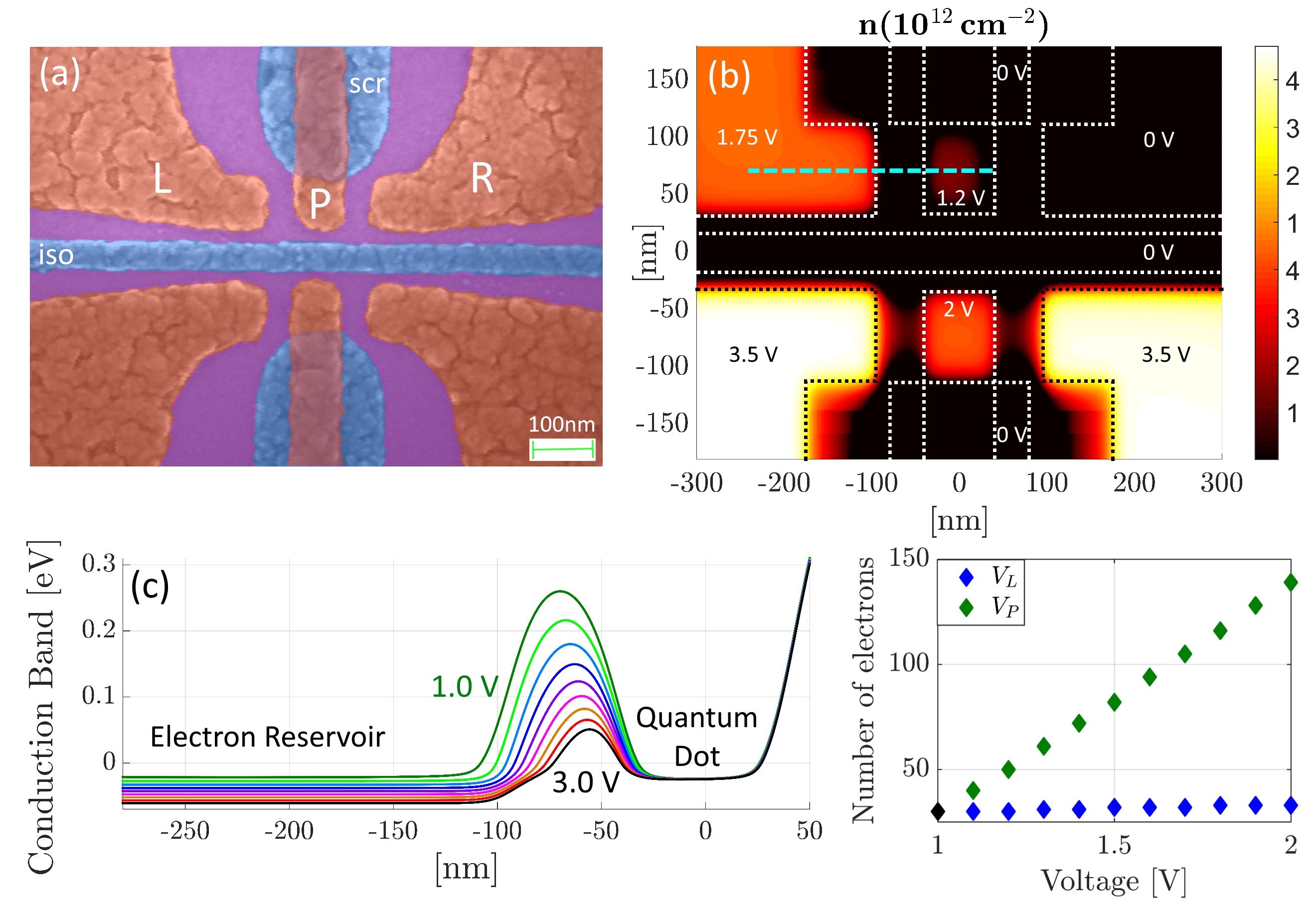}
\caption{\label{fig:SEM}(a) False-colored scanning electron microscope (SEM) image of a device similar to the ones measured. Purple represents the substrate whose top layer is ALD HfO$_2$, while the accumulation and screening metal layers are represented by red and blue, respectively. (b) Using the software package nextnano$^{++}$\cite{Nextnano}, the classical electron density is simulated at low temperature using a finite element Poisson solver for the device geometry denoted by the dotted lines. In the top device, the quantum dot is tuned in the few-electron regime and a single electron reservoir is coupled on the left side. The bottom device functions as a charge sensor and consists of a quantum dot in the many-electron regime with source and drain reservoirs coupled. The voltages applied to each gate are indicated in the figure. (c) 1-D plots of the simulated conduction band of the top device, immediately below the SiO$_2$/Si interface along the dashed horizontal line in (b), for various values of the \textit{L} gate voltage, $V_{L}$, ranging from 1 V to 3 V. The Fermi level is set to 0 eV. (d) Number of electrons in the quantum dot as a function of the gate voltages $V_{L}$ (blue) or $V_{P}$ (green), calculated using a classical simulation in nextnano$^{++}$\cite{Nextnano}. The lever arm ratio of $V_{L}$ over $V_{P}$ is equal to 2.85\%, which shows the decoupling between the voltage $V_{L}$ and the quantum dot occupancy.}
\end{figure}
\begin{figure}
\includegraphics[scale=0.064]{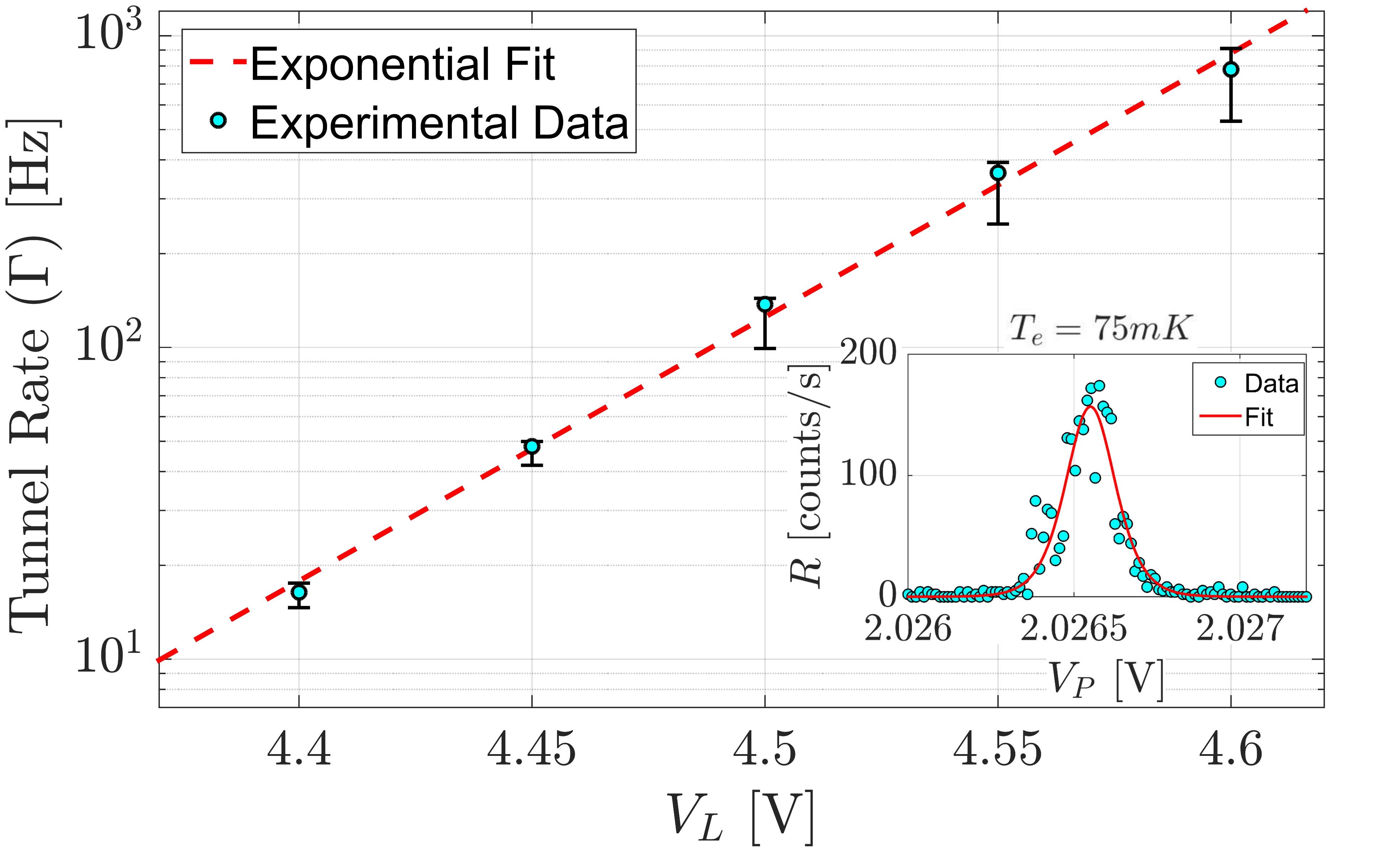}
\caption{\label{fig:Gamma_eCount} Semi-log plot showing the exponential relationship between $\Gamma$ and $V_{L}$, where each experimental data point represents the fitting result between Eq.~\ref{Eq:tunnelrate} and an experimentally determined distribution of $R$ vs. $V_{P}$ (an example $R$ distribution is shown in the inset). The fit shown in the inset gives an estimated electron temperature of 75mK. The exponential fit gives a slope of 8.5 decades/V which provides a measure of the tunability of $\Gamma$ by the reservoir gate voltage $V_{L}$.}
\end{figure}
\begin{figure}
\includegraphics[scale=0.034]{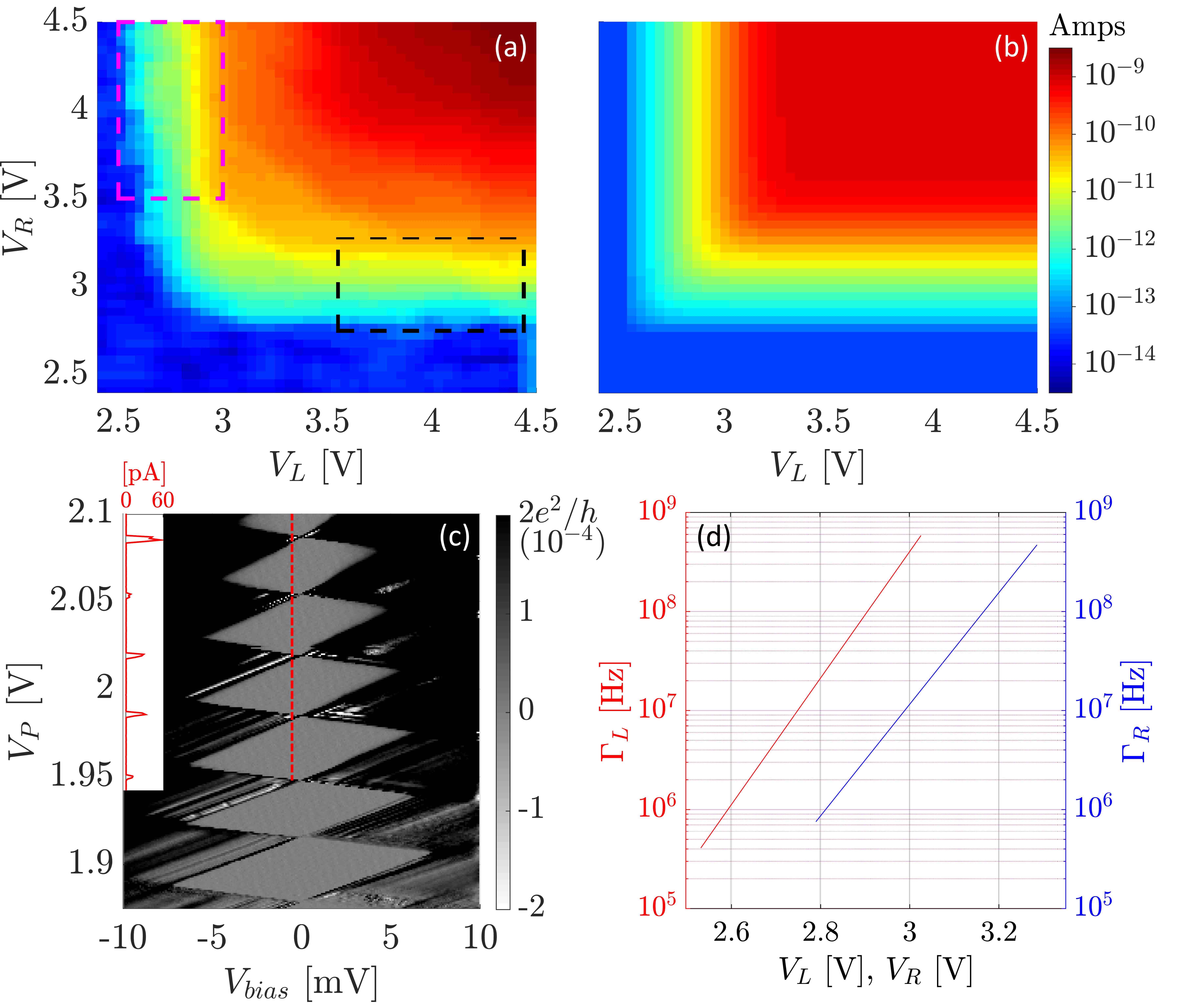}
\caption{\label{fig:Gamma_eTransport}(a) 2D plot of the average experimental current, $I_{avg}$, along the vertical red dashed line in (c) as a function of both $V_{L}$ and $V_{R}$. The coloured dashed boxes enclose regions where either $\Gamma_{L}$ (magenta) or $\Gamma_{R}$ (black) dominates the electron transport. (b) Modeled $I_{avg}$ current based on the approximation that $I_{avg} = e\left(\frac{\Gamma_{L}\cdot\Gamma_{R}}{\Gamma_{L}+\Gamma_{R}}\right)$ and by using a 2D heaviside function to define the pinch-off regions. The colour scale bar for (a) and (b) is in Amperes. (c) Transport measurement of the differential conductance of the target dot showing Coulomb diamonds, which yield an estimate of 185 $\mu$eV/mV for the lever arm of $V_{P}$. (d) Fitting results for the tunability of $\Gamma_{L/R}$ by $V_{L/R}$, respectively, obtained by fitting $I_{avg} = e\Gamma_{L/R}$ to the experimental data within the dashed boxes (a). Both tunnel rates are tuned over roughly three orders of magnitude.}
\end{figure}

Electron or hole spin qubits in silicon quantum dots present a compelling way forward for CMOS-compatible, large-scale quantum information processors. Isotopic removal of nuclear spins enables a dramatic increase in spin coherence times compared to III-V materials,\cite{veldhorst2014addressable,maune2012coherent} and the weakness of the spin-orbit interaction for electrons raises the possibility of coherent spin shuttling.\cite{fujita2017coherent,flentje2017coherent,zhao2018coherent,buonacorsi2018network} A stronger spin-orbit interaction for holes, on the other hand, enables efficient gate-driven single qubit control. \cite{maurand2016cmos} Similar efficient single qubit control has been achieved with electrons using micromagnets to create an artificial spin-orbit field.\cite{takeda2016fault,yoneda2018quantum,kawakami2014electrical,kawakami2016gate} A two-qubit processor has recently demonstrated all the key requirements for computation in one device, namely initialization, implementation of a universal gate set, and readout.\cite{watson2018programmable} In separate devices, single qubit gate fidelities up to 99.96\% \cite{yang2018silicon} and two-qubit (exchange gate) fidelities up to 98\% have been reported,\cite{huang2018fidelity} nearly within reach of expected fault tolerance thresholds for the 2D surface code.\cite{fowler2012surface} Spin qubits have been realized both at the Si/SiO$_2$ interface (MOS qubits) and in Si/SiGe quantum wells.\cite{zajac2015reconfigurable,zajac2018resonantly,takeda2016fault,yoneda2018quantum,kawakami2014electrical,kawakami2016gate} MOS qubits have been realized in a wide range of device geometries, including those fully fabricated in conventional CMOS processing lines.\cite{corna2018electrically,bohuslavskyi2016pauli} Valley degeneracy is one of the key challenges for electron spin qubits in silicon. MOS qubits tend to have, on average, larger valley splitting energies compared to Si/SiGe qubits.\cite{gamble2016valley,rochette2017single,zajac2015reconfigurable} Electric tunability of the valley splitting, to varying degrees, has been demonstrated experimentally \cite{yang2013spin,rochette2017single}.\\
\indent Proposals for scaling up this technology have largely focused on realizing 2D arrays, either compact\cite{veldhorst2017silicon} or in distributed network form,\cite{li2018crossbar,buonacorsi2018network} suitable for surface code implementations. Going beyond a handful of qubits in practice will almost certainly require each qubit to be defined and controlled by as few electrostatic gates as possible, to reduce the number of wires and the density of interconnects. One simplification is to eliminate explicit gates to control dot-reservoir or dot-dot tunnel barriers and instead rely on the device geometry and other nearby gates to tune the tunnel coupling strength. This was demonstrated in MOS devices fabricated by a single gate layer subtractive process \cite{rochette2017single} as well as in a modified CMOS transistor geometry \cite{crippa2017level}. Here, we explore this simplified device concept in an additive two-layer MOS device composed of screening and accumulation gates, in which only 4 gate electrodes define the source, drain and dot and a pair of nanometer size gaps give rise to tunnel barriers. The screening gate in this two-layer metal stack allows for a selective accumulation of electrons without the need for additional confinement depletion gates \cite{rochette2017single}. The accumulation gates form electron reservoirs that define the transport channels coupled to the quantum dot, unlike the device in Ref.~\onlinecite{crippa2017level} where the transport channel is defined via the etching of the Si layer in a silicon-on-insulator structure. The main challenge of the additive two-layer geometry is to demonstrate a robust tunability of the tunnel rate of the tunnel barriers via the voltage control of nearby metal gates. A mirror image device is used as a SET (single-electron transistor) charge sensor. We show that the dot-reservoir tunnel rate, $\Gamma$, can be tuned over $\sim$8 orders of magnitude by the reservoir accumulation gates, which couple only weakly to the dot potential. The device characteristics are clean enough to allow the characterization of the valley-splitting in the few-electron regime by performing magneto-spectroscopy measurements. This work motivates further simplified device geometries, for example, in which the screening gate can be replaced by a thicker dielectric layer so that a quantum dot can be formed by a single metal electrode.\\
\indent Devices are composed of a pair of identical metal-gate defined quantum dots in a mirror image configuration, shown in Fig.~\ref{fig:SEM}a, fabricated on a lightly p-doped (10-20 $\Omega \cdot$ cm) natural Si substrate with 300 nm of thermal SiO$_2$. The first fabrication step defines a device window by wet-etching a region of the SiO$_2$ layer down to around 10 nm thickness. A 6 nm layer of HfO$_2$ is deposited using atomic-layer-deposition (ALD) in order to provide an isolating oxide between n$^+$ implanted regions and the accumulation metal layer. The quantum dot is defined by a two-layer metal gate stack, shown in Fig.~\ref{fig:SEM}a, that is realized using electron-beam (e-beam) lithography and e-beam deposition of aluminum. An oxidation step is performed in between the two metallization steps in order to electrically insulate the metal layers. The first metal layer is referred to as the screening layer and it consists of two screening gates (\textit{scr}) and one isolation gate (\textit{iso}). The purpose of this layer is to fully isolate the top and bottom quantum dots via the \textit{iso} gate, and to prevent accumulation of electrons under the sections of the \textit{P} metal gate that overlap the \textit{scr} gate. The second metal layer is the accumulation layer consisting of two types of gates, and is used to induce electrons at the SiO$_2$/Si interface. The \textit{P} gate defines a single-well potential and controls the electron occupancy of the quantum dot, while the \textit{L} and \textit{R} accumulation gates (i.e. the source/drain accumulation gates) form electron reservoirs to either side of the quantum dots. Ohmic contacts are realized by ion implantation of Phosphorus dopants ($1.5\times10^{15}$ cm$^{-2}$ at 12 keV) more than 100 $\mu$m away from the device region. The source/drain accumulation gates each overlap an ion-implanted area, providing a source of carriers to the device. The completed device undergoes a forming gas (N$_2$ with 5\% H$_2$) anneal at 245$^o$C for 1 hour, with a slow cool down.\\
\indent The control of the dot-reservoir tunnel barriers is commonly assigned to individual metal gates, however in the present design the tunnel barriers are naturally defined by approximately 50 nm wide gaps, illustrated in Fig.~\ref{fig:SEM}a. The barriers are also controlled by the applied potentials on the source/drain accumulation gates (\textit{L} and \textit{R} for the top device). The typical measurement configuration for the pair of quantum dots uses one as the target dot and the other as a SET charge sensor. The charge sensor is tuned in the many-electron occupancy regime and is coupled to the source/drain reservoirs with sufficiently transparent dot-reservoir tunnel barriers to enable direct transport measurements. Meanwhile, the target dot is tuned in the few-electron regime and coupled to only one electron reservoir by a relatively opaque tunnel barrier. A finite element model built using the software package nextnano$^{++}$\cite{Nextnano} solves the Poisson equation and is used to calculate the classical electron sheet density for this device geometry, as shown in Fig.~\ref{fig:SEM}b, where the target dot and charge sensor are at the top and bottom, respectively. The simulated 1D potential profiles shown in Fig.~\ref{fig:SEM}c demonstrate how the applied voltage on the \textit{L} gate, $V_{L}$, controls the size of the dot-reservoir tunnel barrier, while only weakly affecting the electron occupancy of the quantum dot, as shown in Fig.~\ref{fig:SEM}d.\\
\indent The tunability of the dot-reservoir tunnel barrier in the weak coupling regime (i.e. $\Gamma \sim$ 1-1000 Hz) via the applied voltage on the \textit{L} gate, $V_{L}$, is demonstrated by measuring $\Gamma$ at various $V_{L}$ voltages, as shown in Fig.~\ref{fig:Gamma_eCount}. Here, we perform the experiment on a device with a geometry slightly different than the one shown in Fig.~\ref{fig:SEM}a (see Supplementary Material), and the device is tuned into a similar configuration as the one shown in Fig.~\ref{fig:SEM}b, i.e. the target dot is coupled to a single electron reservoir. $\Gamma$ for the target dot is measured by detecting single electron tunnelling events using the charge sensor, while the chemical potential of the target dot, $\mu$, is swept across the Fermi level, $E_{f}$, of its adjacent electron reservoir. The position of $\mu$ with respect to $E_{f}$ is controlled by the \textit{P} gate voltage, $V_{P}$, and the electron tunnelling events are detected by time-resolved measurements of the current through the charge sensor. As $\mu$ moves above or below $E_{f}$, tunnelling is inhibited by a decreasing thermal population of available states. This thermal population follows a Fermi distribution that depends on the effective electron temperature of the electron reservoir. As $\mu$ and $E_{f}$ become aligned, the average number of electron tunnelling events reaches its maximum value and is proportional to the tunnel rate $\Gamma$, which depends on the size of the tunnel barrier. Calculating the number of electron tunnelling events per unit time, $R$, at each value of $V_{P}$ provides a distribution curve for $R$ over a range of $V_{P}$ values. Following the analysis outlined in Ref.~\onlinecite{gustavsson2009electron}, which assumes single-level transport and sequential tunnelling (which are valid in the present experiment), an expression for $R$ is
\begin{eqnarray}
R = f(T_e,E_{f})[1-f(T_e,E_{f})]\Gamma,
\label{Eq:tunnelrate}
\end{eqnarray}
where $f(T_e, E_f)$ is the Fermi distribution and $T_e$ is the effective electron temperature of the reservoir. Every electron tunnelling event causes a sudden change in the measured current of the charge sensor. The procedure of counting electron tunnelling events relies on defining a threshold level, $\delta_{th}$, of the charge sensor current that determines if an electron has tunnelled in/out of the target dot. The details of how $\delta_{th}$ is chosen are discussed in the Supplementary Material. Fig.~\ref{fig:Gamma_eCount} shows the estimated value of $\Gamma$ at different $V_{L}$ values, obtained by fitting Eq.~\ref{Eq:tunnelrate} to the experimentally measured distribution of $R$ at each $V_{L}$, as shown in the inset of Fig.~\ref{fig:Gamma_eCount}, where the fitting parameters are $T_e$, $E_{f}$ and $\Gamma$. The fit shown in the inset of Fig.~\ref{fig:Gamma_eCount} gives an estimated electron temperature equal to 75 mK, while the base temperature of the measurements was 35 mK, as experimentally confirmed by thermometry measurements performed on a GaAs quantum dot device. The major source of error on the estimated $\Gamma$ value comes from the chosen threshold value and not the fitting error in the inset of Fig.~\ref{fig:Gamma_eCount} (see Supplementary Material). This method to estimate $\Gamma$ is limited by the bandwidth of the charge sensor circuit, which in the our experimental setup was about 4 kHz. The major source of uncertainty on the estimated value of $\Gamma$ comes from the chosen value of $\delta_{th}$ since it can significantly affect the maximum value of $R$. This uncertainty is reflected in the error bars shown in Fig.~\ref{fig:Gamma_eCount}. According to Fig.~\ref{fig:Gamma_eCount}, the dependence of $\Gamma$ on $V_{L}$ is well described by an exponential fit, revealing a tunability of about 8.5 decades/V within the range of $V_{L}$ used in the experiment. Note that the particular device used in these experiments had a slightly different gate geometry compared to the device shown in Fig.~\ref{fig:SEM}a (see Supplementary Material).\\
\indent In order to assess the tunability of $\Gamma$ at higher tunnel rates, direct transport measurements were performed on a device nominally identical to the one shown in Fig.~\ref{fig:SEM}a, where the current through the target dot was monitored as a function of the voltages on the \textit{L} and \textit{R} gates, $V_{L}$ and $V_{R}$, respectively. In this configuration, both electron reservoirs are coupled to the target dot. Fig.~\ref{fig:Gamma_eTransport}c shows a charge stability diagram, so-called Coulomb diamonds, measured by direct transport through the target dot, where the lever arm conversion between $V_{P}$ and the applied bias voltage, $V_{bias}$, is 185 $\mu$eV/mV. At constant values for both $V_{L}$ and $V_{R}$, the average current ($I_{avg}$) over a sweep of $V_{P}$ was obtained along the dashed red line ($V_{bias}$ = 0.5 mV) in Fig.~\ref{fig:Gamma_eTransport}c, which encompasses at least four current peaks. The sweep of $V_P$ was repeated while $V_{L}$ and $V_{R}$ were each separately stepped from 2.4 V to 4.5 V. Fig.~\ref{fig:Gamma_eTransport}a shows $I_{avg}$ as a function of $V_{L}$ and $V_{R}$, where the current is seen to pinch off at $V_{L} = $2.57 V and $V_{R} = $2.79 V. The transport current can be approximated by the expression $I_{avg} = e\left(\frac{\Gamma_{L}\cdot\Gamma_{R}}{\Gamma_{L}+\Gamma_{R}}\right)$, where $e$ is the electron charge and $\Gamma_{L/R}$ represent the tunnel rates between the dot and the left/right reservoir, respectively. Based on the result shown in Fig.~\ref{fig:Gamma_eCount}, $\Gamma_{L/R}$ is assigned an exponential relationship with respect to $V_{L/R}$ in the form of $\Gamma_{L/R} = e^{[a_{L/R}(V_{L/R} + b_{L/R})]}$, where $a_{L/R}$ and $b_{L/R}$ are fitting parameters. Inside the dashed boxes shown in Fig.~\ref{fig:Gamma_eTransport}a, only one tunnel coupling dominates the electron transport, and the expression for $I_{avg}$ can be simplified to $I_{avg} = e\Gamma$ in those regions, where $\Gamma$ represents the dominant tunnel rate. Fitting the experimental data in each dashed box yields a tunability of 6.4 decades/V and 5.6 decades/V for $\Gamma_{L}$ and $\Gamma_{R}$, respectively, over a tunnel rate range of $\sim 10^5 - 10^9$ Hz, as seen in Fig.~\ref{fig:Gamma_eTransport}d. Using the fitting results for the individual $\Gamma_{L}$ and $\Gamma_{R}$ and the expression for $I_{avg}$, an approximate model of the transport current can be calculated over the whole $V_{L}$ and $V_{R}$ parameter space to obtain the 2D plot shown in Fig.~\ref{fig:Gamma_eTransport}b, where the pinch off regions are enforced via a 2D heaviside function and a maximum current is imposed to resemble the saturation behaviour in the upper right corner of the measured current in Fig.~\ref{fig:Gamma_eTransport}a. The current saturation is due to the minimum resistance of the device channel in the high-accumulation regime (see Supplementary Material) and it lies outside both dashed boxes in Fig.~\ref{fig:Gamma_eTransport}a, hence it does not affect the fitting results shown in Fig.~\ref{fig:Gamma_eTransport}d. The experimental data in Fig.~\ref{fig:Gamma_eTransport}a was also used to determine the coupling strength between each \textit{L} and \textit{R} gate and $\mu$ in the target dot, in order to obtain a lever arm of about 9.0 $\mu$eV/mV for each $V_{L}$ and $V_{R}$. This lever arm is slightly less than $5\%$ of the lever arm of $V_{P}$, which demonstrates a good decoupling between the tunability of $\Gamma$ and the dot potential and it agrees with the value obtained in the classical simulation shown in Fig.~\ref{fig:SEM}. Overall, the tunability of $\Gamma$ at low and high tunnel rates was similar in spite of the small differences in device geometries utilized in each experiment (Fig.~\ref{fig:Gamma_eCount} and Fig.~\ref{fig:Gamma_eTransport}). It is worth mentioning that the device geometry used for the electron counting experiment could also be tuned such that its tunnel rates were high enough and would enable direct transport measurements (see Supplementary Material), therefore this device geometry could also have been utilized for similar direct transport experiments shown in Fig.~\ref{fig:Gamma_eTransport}. \\
\begin{figure}
\includegraphics[scale=0.0438]{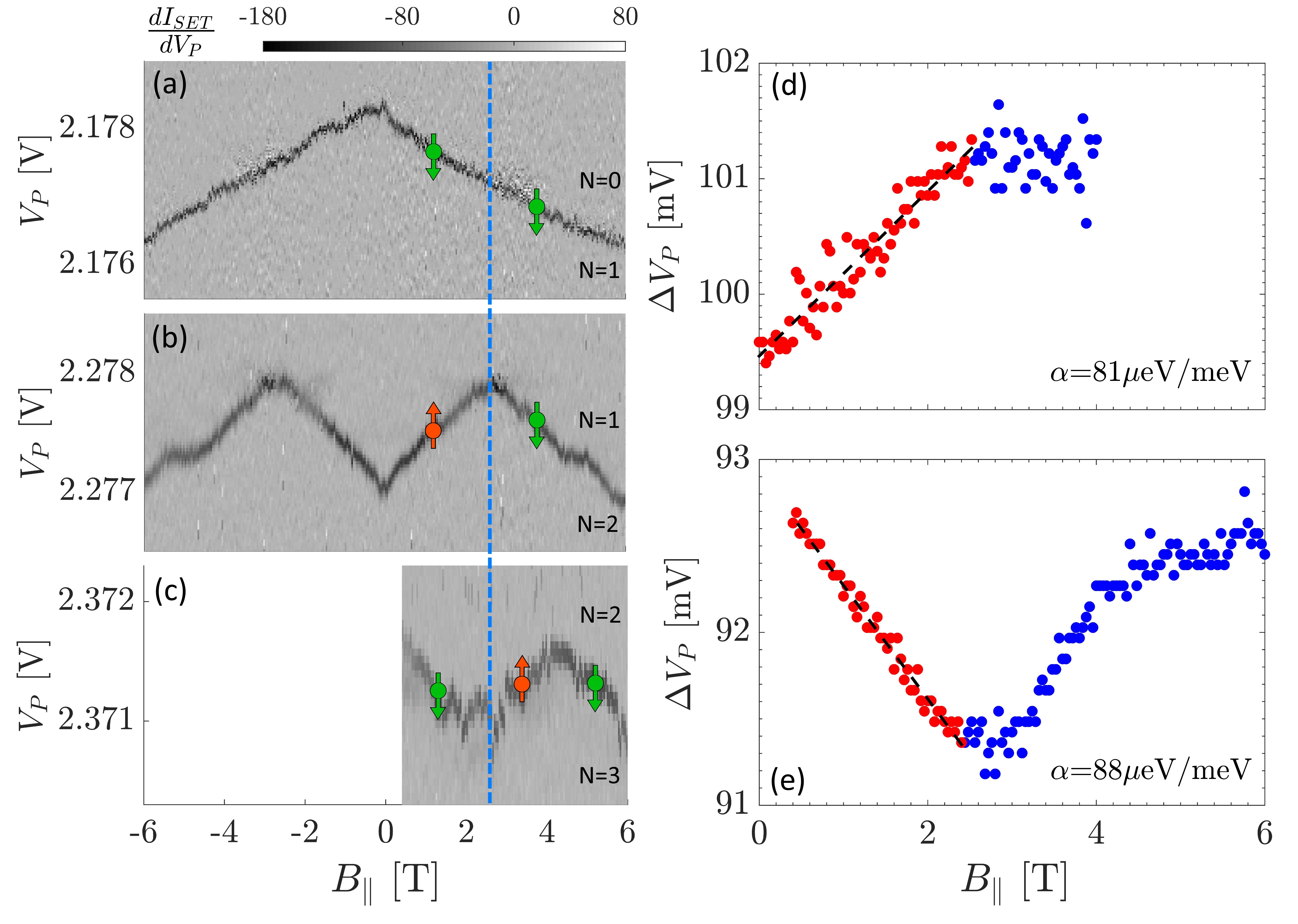}
\caption{\label{fig:ValleySplit} Magneto-spectroscopy results acquired by charge sensing, showing spin filling of the target dot for the (a) $N=0\rightarrow1$, (b) $N=1\rightarrow2$, and (c) $N=2\rightarrow3$ electron transitions with an in-plane magnetic field $-6$ T $<B_{\parallel}< 6$ T. The spin state is denoted by the green and red arrow symbols. A spin-state crossover is evidenced by the kink feature in (a) and (b) occurring at 2.5 T (denoted by vertical dashed line), which corresponds to a valley splitting $\Delta E_{V} \approx$ 290 $\mu$eV. (d) Difference between the $V_P$ values of the $N=0\rightarrow1$ and $N=1\rightarrow2$ charge transitions. The linear slope before the spin-state crossover indicates opposite spins, while the flat region indicates the same spin. The lever arm for $V_P$, $\alpha$, is calculated assuming an electronic $g$-factor of 2. (e) The difference between the $V_P$ values of the $N=1\rightarrow2$ and $N=2\rightarrow3$ charge transitions.}
\end{figure}
\indent The lifting of the energy degeneracy between the two low-lying valley states, $E^{-}_{V}$ (ground) and $E^{+}_{V}$ (excited), due to the electronic confinement along the z-axis at the SiO$_2$/Si interface,\cite{zwanenburg2013silicon, friesen2010theory} has an important role in ensuring that a spin qubit remains coherent. An insufficient valley splitting can provide a spin-flip mechanism that can lift Pauli spin blockade and prevent the use of spin-to-charge conversion techniques, which is key in single-shot spin readout.\cite{Tagliaferri2018} For these reasons, the magnitude of the valley splitting energy, $\Delta E_{V}$, is investigated by a magneto-spectroscopy technique, where an applied in-plane magnetic field, $B_{\parallel}$, lifts the spin degeneracy of $E^{-}_{V}$ and $E^{+}_{V}$ and leads to a particular spin filling pattern for the dot. The device geometry used for this magneto-spectroscopy experiment was the same as the one used for the electron counting experiments (see Supplementary Material). The lowest four available spin-valley states are $E^{-\downarrow}_{V}$, $E^{-\uparrow}_{V}$, $E^{+\downarrow}_{V}$ and $E^{+\uparrow}_{V}$, where $\uparrow$ and $\downarrow$ define the spin state. Spin filling of the target dot is studied by sweeping $V_{P}$ such that an electron from a tunnel coupled reservoir can tunnel into the target dot and occupy the lowest available energy state. In this case, $\Gamma$ is much larger than the sweep rate of $V_{P}$ and the electron tunnelling event is detected with the aid of a charge sensor. The $V_{P}$ sweep is repeated as $B_{\parallel}$ is varied within a range of $\pm$ 6 T. A plot of the charge sensor signal vs. $B_{\parallel}$ is shown in Fig.~\ref{fig:ValleySplit}, where changes in the dot chemical potential, $\mu$, are tracked in the few electron regime as $B_{\parallel}$ varies. Fig.~\ref{fig:ValleySplit}a-c tracks the chemical potential $\mu(N)$ for the charge transitions $N=0\rightarrow1$, $N=1\rightarrow2$, and $N=2\rightarrow3$, respectively. In Fig.~\ref{fig:ValleySplit}a, the $N=0\rightarrow1$ transition involves an electron filling the $E^{-\downarrow}_{V}$ state, where $\mu(N)$ decreases linearly with increasing $B_{\parallel}$. In the $N=1\rightarrow2$ transition, the second electron initially occupies the $E^{-\uparrow}_{V}$ state for $|B_{\parallel}|<$2.5 T, and subsequently favours occupying the $E^{+\downarrow}_{V}$ state for $|B_{\parallel}|>$2.5 T. This spin flip is due to the crossing of the $E^{-\uparrow}_{V}$ and $E^{+\downarrow}_{V}$ states that occurs when $g\mu_B B_{\parallel}=\Delta E_{V}$, as indicated by the kink seen in Fig.~\ref{fig:ValleySplit}b at $B_{\parallel}=$ 2.5 T. This gives an estimate for $\Delta E_{V}$ of 290 $\mu$eV. A kink is also observed at the same magnetic field for the $N=2\rightarrow3$ transition (note the signal in Fig.~\ref{fig:ValleySplit}c appears noisier because the scan was acquired at a lower resolution). In Figures~\ref{fig:ValleySplit}d and \ref{fig:ValleySplit}e show the difference in $V_{P}$ values between adjacent transitions, $\Delta V_{P}$, corresponding to the electron addition energies. Assuming an electronic $g$-factor equal to 2, lever arms for $V_{P}$ are estimated to be $81~\mu$eV/mV and $88~\mu$eV/mV for the two transitions.\\
\indent A second spin flip is observed in Fig.~\ref{fig:ValleySplit}c at $B_{\parallel}\approx$ 4.3 T which could be due to the $\downarrow$ state of an excited orbital. This would suggest an orbital energy spacing of about 500 $\mu$eV. Similar spin flip features in the few-electron regime are reported in Ref.~\onlinecite{lim2011spin}, which are partially explained by the mixing of valley and orbital states when $\Delta E_{V}$ is comparable to the orbital energy, $E_{orb}$. An estimate for $E_{orb}$ can be obtained\cite{kouwenhoven1997electron} using $E_{orb}=\frac{2\pi\hbar^2}{g_s g_v m^*A}$, where $m^*$ is the transversal effective mass of the electron, $A$ is the area of the quantum dot, and $g_v$ and $g_s$ are the valley and spin degeneracy, respectively. In the case of non-degenerate valley states, $g_v=1$ and $g_s=2$. Approximating the quantum dot as a thin disk, the radius of the quantum dot equals $r=\frac{e^2}{8\epsilon_o\epsilon_r E_c}$, where $E_C$ is the charging energy of the quantum dot. $E_C$ is estimated to be 8.2 meV using the addition voltage and the approximate lever arm (88 $\mu$eV/mV) shown in Fig.~\ref{fig:ValleySplit}e. Based on these values, $E_{orb}\approx$ 680 $\mu$eV which is $\approx$ 1.4 times larger than the observed energy (500 $\mu$eV) corresponding to the second kink in Fig.~\ref{fig:ValleySplit}c, which suggests that valley-orbital mixing may play a role here. Nonetheless, the quantity of most interest for spin qubits is the aforementioned "ground-state" gap of 290 $\mu$eV. A subtle feature we cannot explain is observed in Fig.~\ref{fig:ValleySplit}b at approximately $\pm$5 T, a small region where the slope is close to zero. This is not explained based on the simple spin filling model for the $N=1\rightarrow2$ transition, and it cannot be due to a lower energy $\uparrow$ state since no corresponding feature is seen for the $N=0\rightarrow1$ transition. \\
\begin{figure}
\includegraphics[scale=0.205]{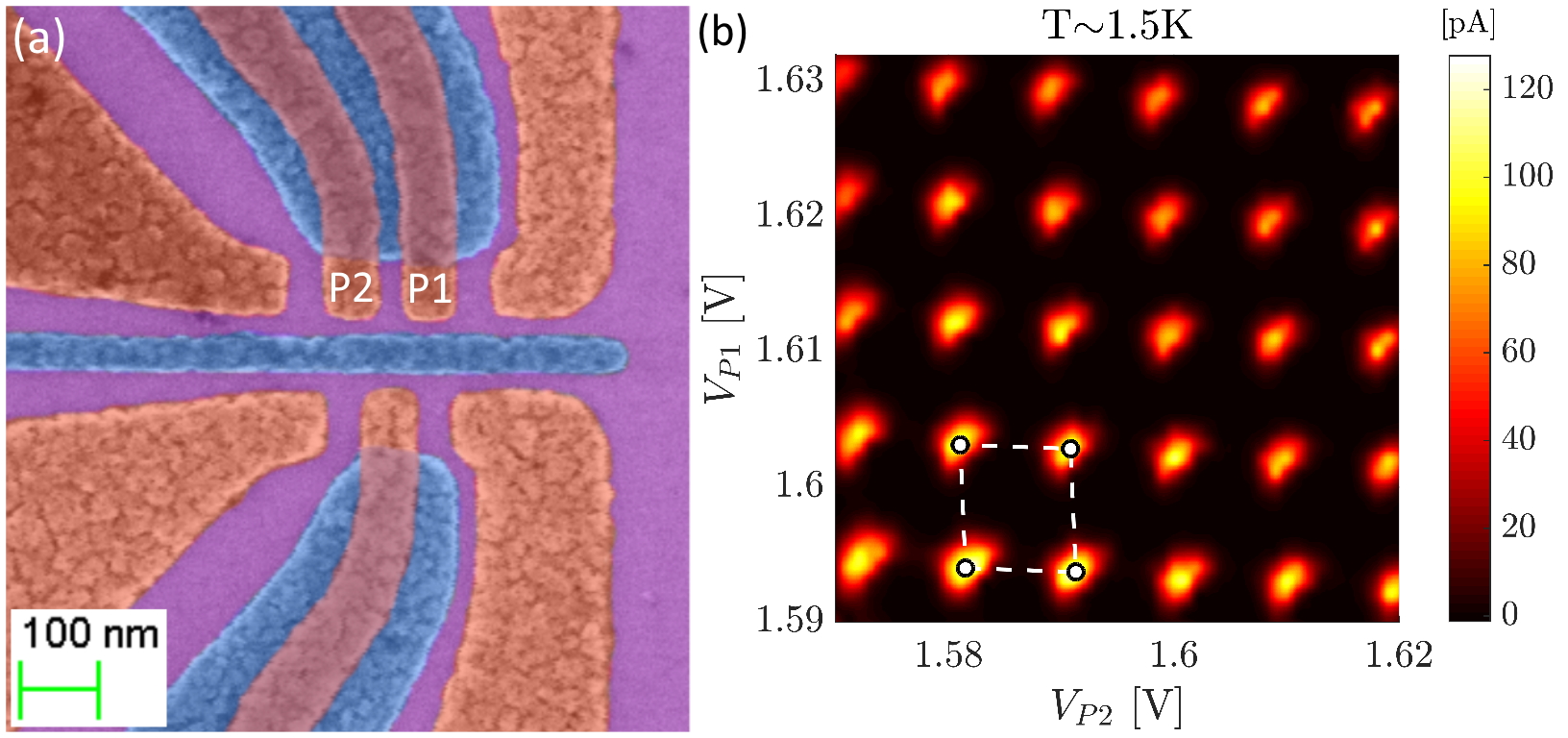}
\caption{\label{fig:dqd} Panel (a) shows a false-coloured SEM image of a double dot device with plunger gates labelled P1 and P2 corresponding to dots 1 and 2, respectively. Direct transport in the many electron regime measured at T$=1.5$ K is shown in panel (b), where the plunger gate voltages are varied at a fixed source-drain bias of 1 mV. The four dashed lines and dots indicate fitting to a constant-interaction capacitance model, from which we extract the direct gate capacitances $16.9\pm0.6$ aF (P1-dot1) and $17.5\pm0.2$ aF (P2-dot2), and both cross capacitances $\approx 0.8$ aF.}
\end{figure}
\indent This device geometry can be extended to a linear array of multiple quantum dots in series. In figure~\ref{fig:dqd} we show direct transport characterization of a double quantum dot device measured at T$\approx1.5$ K. Fig.~\ref{fig:dqd}a is an SEM image of a nominally identical device, with the plunger gates of the double quantum dot labeled P1 and P2. A section of the charge stability diagram in the many electron regime is shown in Fig.~\ref{fig:dqd}b, indicating a well-defined double quantum dot. Direct and cross gate capacitances were determined by fitting the relative positions of the current peaks to a capacitance model \cite{Penfold2017} (fit shown by dashed lines). The direct gate-dot capacitances for P1 and P2 were $16.9\pm0.6$ aF and $17.5\pm0.2$ aF, respectively. The cross-capacitances were $0.8\pm0.4$ aF (P1 to dot 2) and $0.8\pm0.1$ aF (P2 to dot 1). We measured these values across three nominally identical devices and found that the P2 direct capacitances were systematically larger, with an average of 18.3 aF compared to 16.8 aF for P1. The device-to-device variation was low: within 2$\%$ for the P1 capacitances and within 8$\%$ for the P2 capacitances. Clearly, more devices should be characterized before drawing statistical conclusions, but these preliminary results suggest that our design and fabrication methods yield good reproducibility. We expect that a linear array of dots with this minimal gate layout will be useful for charge/spin shuttling \cite{mills2019shuttling}, a key ingredient in some proposals for a scalable spin qubit processor in silicon \cite{buonacorsi2018network, li2018crossbar}. Explicit tunnel gate electrodes will probably still be needed for fine multi-qubit control, e.g. two-qubit exchange gates and other qubit operations. \\
\indent In conclusion, the device geometry demonstrated here presents a simplification to the usual metal-gate stack used in Si MOS quantum dots by removing tunnel barrier gates and relying on a reservoir accumulation gate for control over the dot-reservoir tunnel coupling. It was shown that the magnitude of $\Gamma$ can be controlled with a tunability of up to 8.5 decades/V, while maintaining good decoupling between the accumulation gate and the chemical potential of the dot. This geometry is useful for charge sensors that are robust and easy to tune up. Furthermore, magneto-spectroscopy experiments enabled by charge sensing demonstrate that the device characteristics are clean enough to perform spin-filling measurements in the few-electron regime, where a ground state gap of 290 $\mu$eV was observed. We also demonstrated the extension of this geometry to a double quantum dot, and suggested that this could be further extended to linear dot arrays ideal for electron shuttling experiments. Further simplifications can be pursued, for example, replacing the screening gate layer by a suitably thick dielectric so that a quantum dot can be defined by a single gate electrode. Such device simplifications benefit the scalability prospects of Si MOS quantum dots as candidates for realizing spin-based quantum processors.
\section*{Supplementary Material}
A supplementary file includes device details, analysis methods for tunnel rates using both counting statistics and direct current, and fitting methods for the double dot data.  
\begin{acknowledgments}
This research was undertaken thanks in part to funding from the Canada First Research Excellence Fund and NSERC. We thank the staff at the Quantum NanoFab Facility at the University of Waterloo for technical support in the fabrication of devices. We acknowledge Kyle Willick for technical help and Brandon Buonacorsi for helpful discussions. EBR acknowledges a Nanofellowship sponsored by the Waterloo Institute for Nanotechnology. 
\end{acknowledgments}

\providecommand{\noopsort}[1]{}\providecommand{\singleletter}[1]{#1}%
%

\newpage

\section{Supplementary Material for\\ ``Few-electrode design for silicon MOS quantum dots''}

\begin{figure}[h!]
\centering
\includegraphics[scale=0.28]{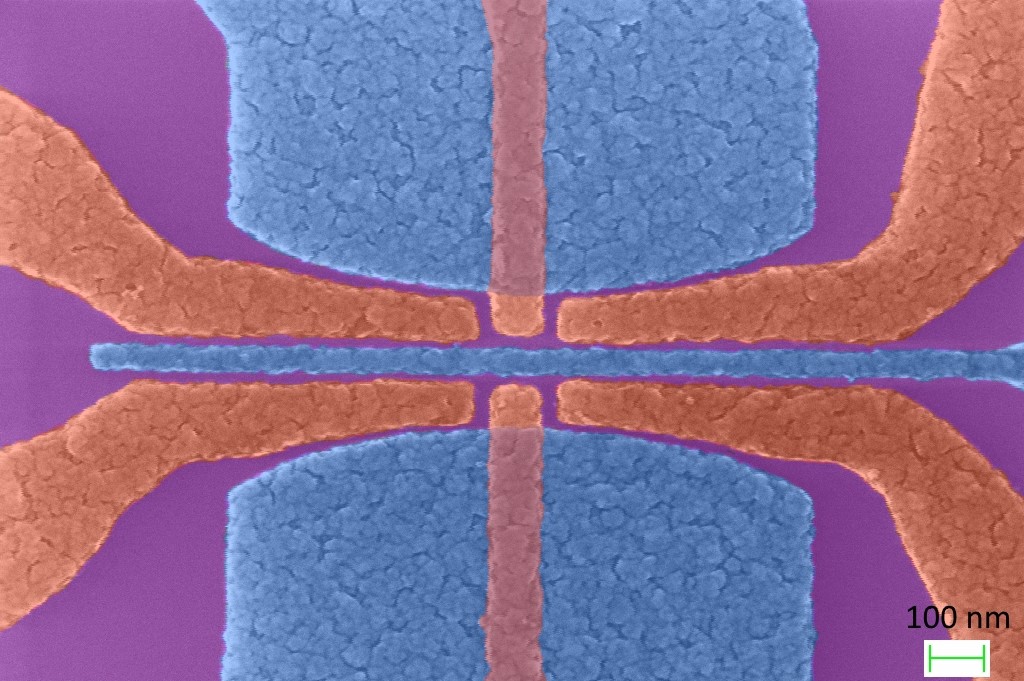}
\caption{\label{fig:S1} False-coloured scanning electron microscope (SEM) image of a device nominally identical to the ones measured in the magneto-spectroscopy and electron counting experiments discussed in the main text.}
\end{figure}

\subsection{Device geometry}
The data shown in Fig. 2 and Fig. 4 in the main text was obtained by measuring a device geometry nominally identical to the one shown in Fig.~\ref{fig:S1}, which differs from the SEM image shown in Fig. 1a of the main text. The main difference is an elongation of the accumulation gates used for the formation of electron reservoirs and a broadening of the screening gates. All other fabrication steps are identical to the ones outlined in the main text.

\begin{figure}[h!]
\centering
\includegraphics[scale=0.06]{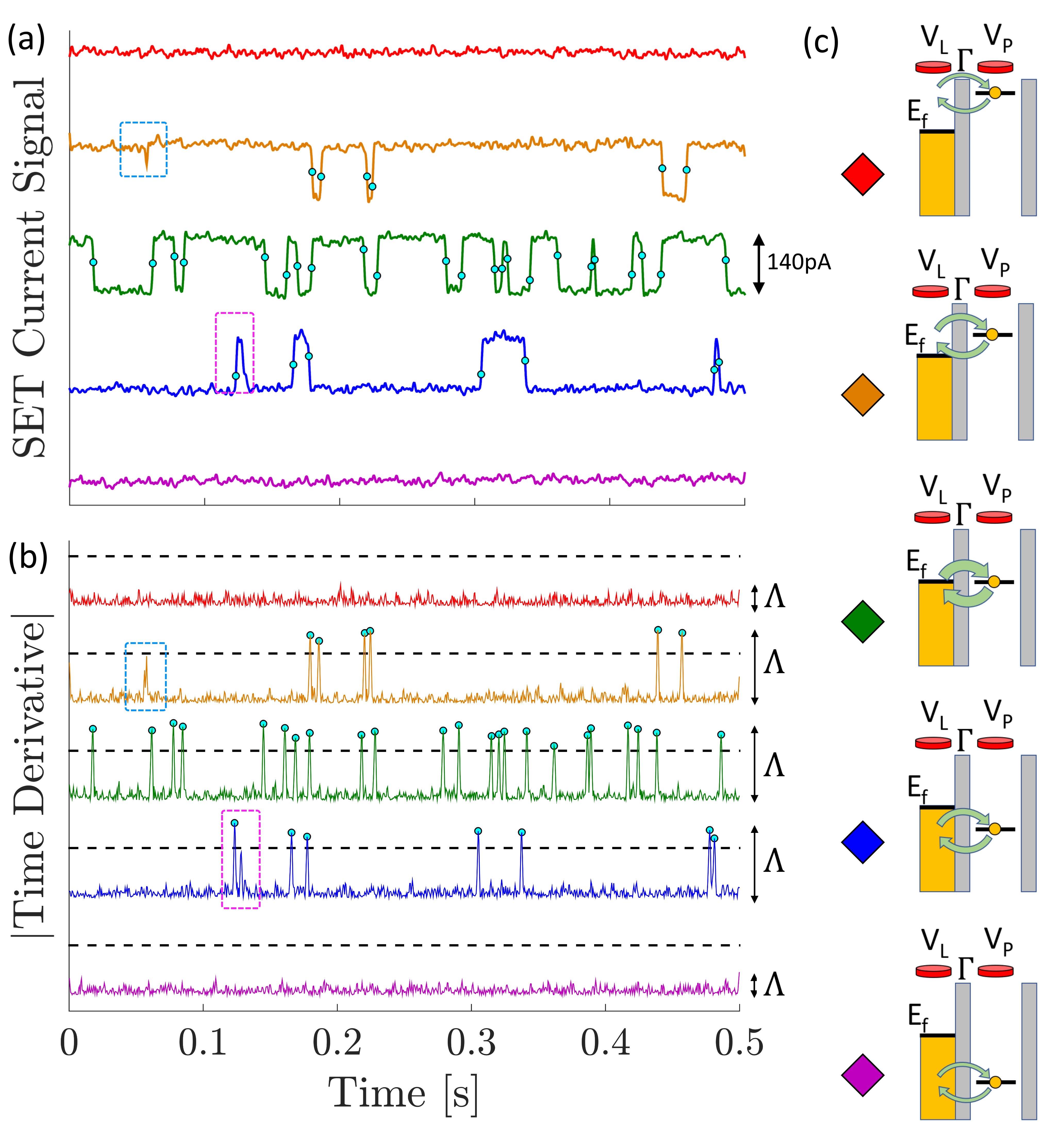}
\caption{\label{fig:S2}(a) Time-resolved traces of the current through the SET charge sensor at various values of $V_P$ (colour coded). Clear electron tunnelling events are observed with an average frequency that varies as $V_{P}$ is changed. (b) Time derivative of the traces shown in (a) where electron transitions are indicated by peaks. The total number of electron transitions at each $V_P$ is determined as the number of peaks that are above a chosen threshold value, denoted by the horizontal black dashed line. (c) Schematic illustrations of the alignment between the chemical potential of the dot and the Fermi level of the lead, which are colour coded to match the traces shown in (a) and (b).}
\end{figure}

\begin{figure*}[h!]
\includegraphics[width=0.8\textwidth]{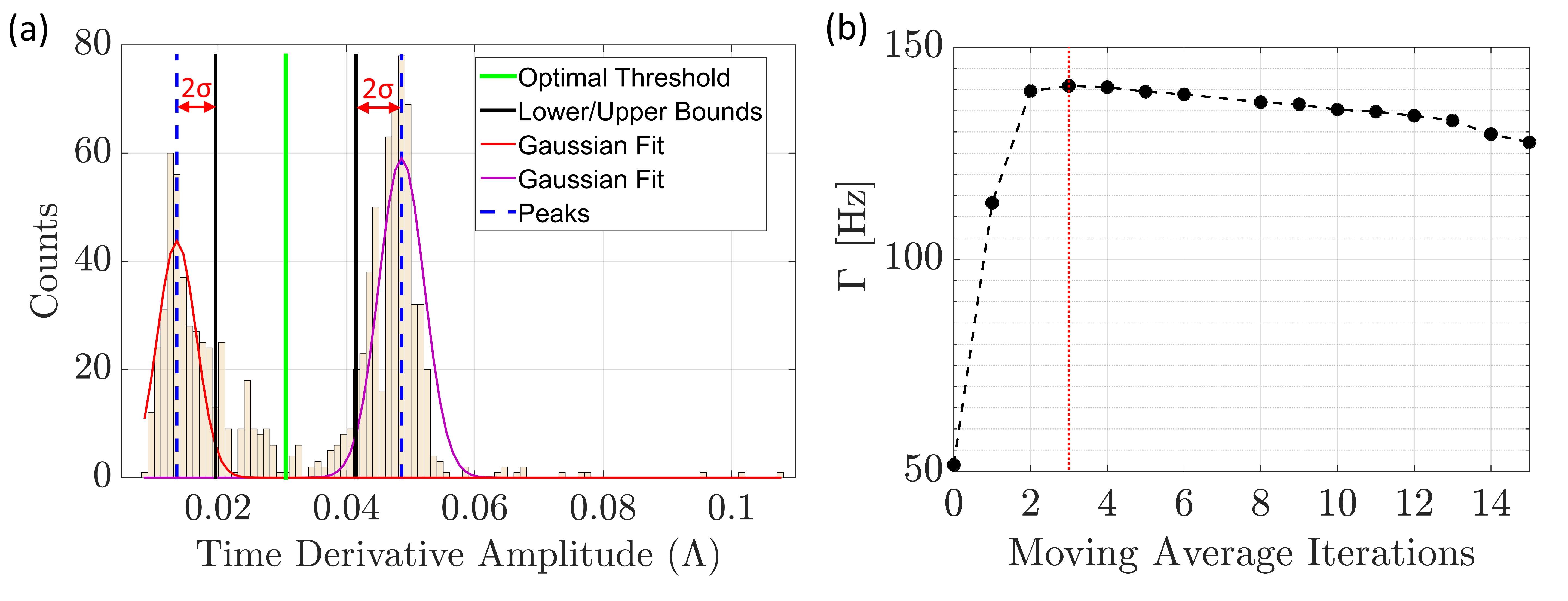}
\caption{\label{fig:S3}(a) Histogram of the time derivative amplitude $\Lambda$ (as illustrated in Fig.~\ref{fig:S2}b) where two clear peaks are observed (dashed blue) and fit to Gaussian curves (solid red). The peak on the right represents the distribution of $\Lambda$ values that correspond to electron tunnelling events. The peak on the left represents the distribution of $\Lambda$ values corresponding to the noise floor of the SET current signal. The bounds of the threshold value (solid black) are placed at two standard deviations from each peak while the optimal threshold value (solid green) is set at the midpoint between the these two bounds. (b) A plot of the estimated $\Gamma$ as a function of the number of applied iterations of a 3-point moving average on the time-resolved experimental data. The $\Gamma$ increases significantly for the first few data points since averaging improves the signal-to-noise ratio. Beyond an optimal value of 3 iterations, $\Gamma$ slowly decreases, since excessive averaging compromises the detection of tunnelling events.}
\end{figure*}

\begin{figure*}[h!]
\includegraphics[width=0.8\textwidth]{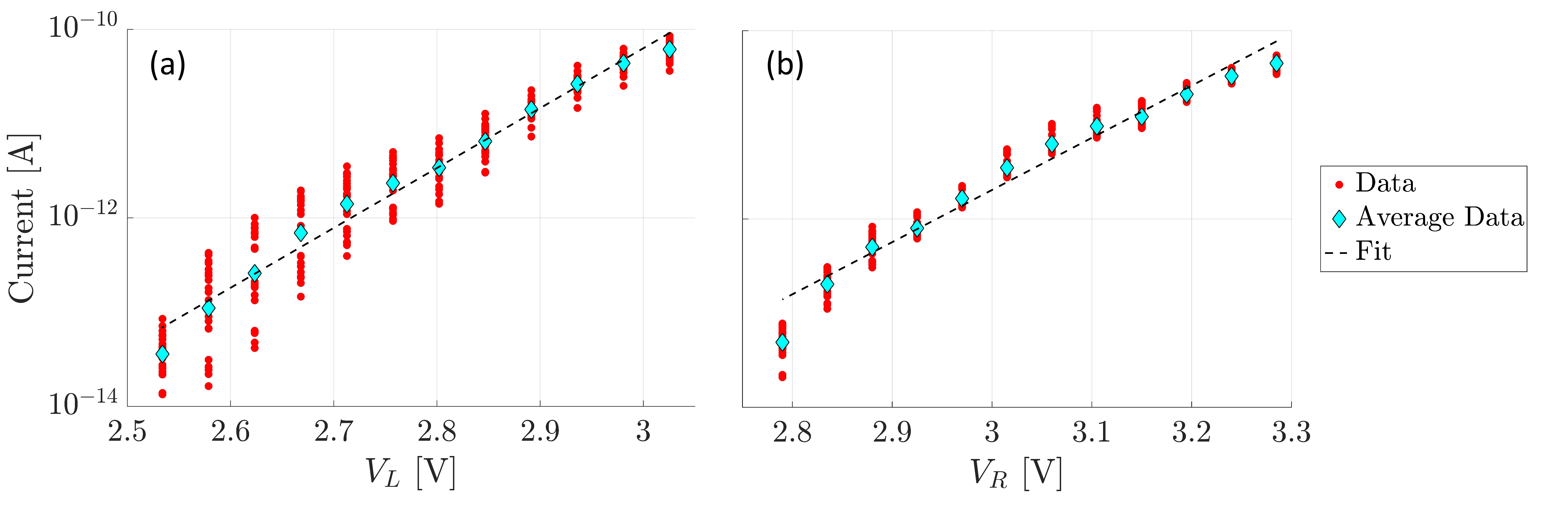}
\caption{\label{fig:S4}(a) The exponential fit (dashed black line) to the experimentally measured average current $I_{avg}$ in the case where $\Gamma_L$ dominates transport. (b) Similar to (a), but for the case where $\Gamma_R$ dominates. The experimental data is shown by the red circles and the average of the data is shown by the cyan diamonds. The corresponding voltage dependencies for $\Gamma_L$ and $\Gamma_R$ are extracted from (a) and (b) and are shown in Fig. 3d in the main text.}
\end{figure*}

\subsection{Counting electron tunnelling events}
In the data shown in Fig. 2 of the main text, each data point was obtained through the analysis of a set of time-resolved measurements for the current of the SET charge sensor at varying values of $V_{P}$. An example of these types of time-resolved measurements is shown in Fig.~\ref{fig:S2}a, where each colored trace corresponds to a different value of $V_{P}$ and the number of electron tunnelling events increases as the chemical potential of the dot aligns to the Fermi level of the reservoir, as illustrated in Fig.~\ref{fig:S2}c (colour coded). The absolute value of the time-derivative of these colored traces is shown in Fig.~\ref{fig:S2}b, where the electron tunnelling events are seen as sharp peaks in the time-derivative signal. Individual electron tunnelling events can be counted by determining the number of peaks that are above a common threshold value, shown by the black dashed horizontal lines in Fig.~\ref{fig:S2}b, for all the time-derivative signals. The method for choosing an optimal threshold value begins by plotting a histogram of the maximum amplitude of each time derivative signal referred to as $\Lambda$ and shown in Fig.~\ref{fig:S2}b by the vertical double-head arrows. This histogram is shown in Fig.~\ref{fig:S3}a, where there are two clear peaks which are fit to a Gaussian curve. The peak centered at the higher value of $\Lambda$ represents most of the electron tunnelling events, while the other peak at the lower value of $\Lambda$ corresponds to an absence of an electron tunnelling event (signal noise floor). The lower and upper bounds for the threshold value are placed two standard deviations away from the each peak, as shown in Fig.~\ref{fig:S3}a. These bounds are used as the upper and lower error bars in Fig. 2 of the main text. The optimal threshold value is set to the midpoint between these two bounds.\\
\indent The purple dashed rectangle in Fig.~\ref{fig:S2}a shows an electron tunnelling out and back into the quantum dot, which matches to the two peaks shown inside the purple dashed rectangle in Fig.~\ref{fig:S2}b. The peak corresponding to the electron tunnelling back into the dot is not properly captured as an electron tunnelling event since the corresponding peak in Fig.~\ref{fig:S2}b lies below the threshold value. This could be corrected by choosing a lower threshold value, however this new threshold value would have also counted the peak shown in the cyan dashed rectangle in Fig.~\ref{fig:S2}b as an electron tunnelling event, even though it clearly is not, as seen in Fig.~\ref{fig:S2}a. This highlights the importance of determining an optimal threshold value to ensure that the extracted tunnel rate, $\Gamma$, is as accurate as possible.\\
\indent The electron counting analysis is performed after a 3-point moving average is repeatedly applied on the SET current signal, which helps to reduce noise and improve the accuracy of the estimated $\Gamma$. Fig.~\ref{fig:S3}b shows the estimated value of $\Gamma$ as a function of the number of times that the moving average was applied on the data, where the optimal value of $\Gamma$ is chosen at the peak of the curve.\\
\subsection{Effective device resistance in transport measurements}
The 2D plot shown in Fig. 3b in the main text was calculated by assuming that the total resistance of the quantum dot device, $R_{tot}$, was equal to $R_{dot}+R_{min}$, where $R_{dot}$ is the equivalent resistance for electron transport through the two tunnel barriers and $R_{min}$ is the minimum resistance for the channel formed by the accumulation gates. $R_{dot}$ is given by $\frac{V_{bias}}{I_{dot}}$ where $I_{dot} = e\left(\frac{\Gamma_{L}\cdot\Gamma_{R}}{\Gamma_{L}+\Gamma_{R}}\right)\cdot H(V'_{L},V'_{R}) + I_o\cdot\left[1 - H(V'_{L},V'_{R})\right]$. $H(V'_{L},V'_{R})$ is a 2D Heaviside function which enforces pinch-off regions below the pinch-off voltages $V'_{L}$ and $V'_{R}$, while $I_o$ is the minimum experimentally measurable current. Therefore, $R_{dot}$ depends on the value for $V_L$ and $V_R$ due to the voltage dependence of $\Gamma_L$ and $\Gamma_R$. $R_{min}$, $I_o$, $V'_{L}$ and $V'_{R}$ are all fitting parameters. The total current through the dot is then calculated as $I_{tot}=\frac{V_{bias}}{R_{tot}}$. Figure~\ref{fig:S4} shows the raw data underlying the fits shown in Fig. 3d of the main paper. Since these fits are done near the pinch-off regions, the total device resistance is dominated by the dot resistance and the channel resistance can be ignored.  \\
\subsection{Fitting double dot data}
\indent Beginning with a charge stability diagram as shown in Fig. 5(b) of the main paper, a home-written code first sets a minimum current threshold and only data above the threshold is kept, defining the current regions around the triple points. Next, a cluster-scan algorithm called DBSCAN is used to identify these regions of good signal separately as clusters and number them. Then, a set of four neighboring clusters that have the highest current signal are identified. Their centroids are calculated and a parallelogram is fit to the centroids. The dimensions of the parallelogram give the gate and cross-gate capacitances based on the constant-interaction model (see Appendix B in Ref. 33 of the main text). The main source of error comes from setting a current threshold value to identify current regions around triple points. This error propagates to the final capacitance values via the centroid calculations. To estimate errors in final capacitance values, the code was iterated over a range of current threshold values chosen by visual examination. The standard deviation in the capacitance values across these iterations is reported as the error.

\end{document}